# Structural Phase Transitions and Vertical Mode Spectra in 2D Finite Plasma Crystals

Ke Qiao and Truell W. Hyde

*Abstract*—**A numerical simulation utilizing a box_tree code is used to investigate the structure and vertical mode spectrum of finite two-dimensional (2D) plasma crystals. The overall structural symmetry of the system is examined for various Debye lengths and a transition from a predominantly hexagonal structure to a structure having concentric rings along the outer edge and hexagonal lattice symmetry in the interior is shown to develop as the Debye length increases. Both the vertical and horizontal oscillation modes for this type of system are investigated where the horizontal mode spectra is shown to agree with published results while the vertical mode spectra obtained is shown to agree with an independent analytical method. The fundamental frequency for the vertical modes decreases as the mode number $l$ decreases and is shown to have a maximum corresponding to that which would exist if the system acted *in toto* as a solid plane. For low frequency vertical modes, the largest amplitude particle motion is concentrated within a few inner rings with the outer rings remaining almost motionless. Both of these are in direct contrast to the data obtained for the horizontal modes where it is shown that at high frequencies the largest amplitude particle motion is concentrated within the inner rings.**

## I. INTRODUCTION

COMPLEX (dusty) plasma has been a rapidly expanding research field since the discovery of plasma crystals in 1994 [1]-[3], primarily due to the important role such plasmas play across fields as disparate as astrophysics and nanofabrication. The majority of plasma crystals investigated in laboratories on earth consist of two-dimensional (2D) or quasi-2D systems of dust particles levitated within a rf plasma sheath above a powered lower electrode. Such systems generally attain equilibrium vertically through attaining a balance between the gravitational and electrical forces within the sheath, and are confined on the horizontal direction through either a curved lower electrode or a ring placed on the lower electrode. The resulting potential well confining the particles on both the vertical [4] and horizontal [5], [6]

Manuscript received September 27, 2007. This work was supported in part by the National Science Foundation and the Department of Education.

Ke Qiao is with the Center for Astrophysics, Space Physics & Engineering Research, One Bear Place 97310, Baylor University, Waco, TX, USA. (e-mail: ke_qiao@ baylor.edu).

Truell W. Hyde is with Center for Astrophysics, Space Physics & Engineering Research, One Bear Place 97310, Baylor University, Waco, TX, USA. (phone: 254-710-3763; fax: 254-710-7309; e-mail: truell_hyde@baylor.edu).

directions is known to be parabolic to at least first order with the interaction between particles given by a Yukawa potential of the form

$$U(r) = q \exp(-r / \lambda_D) / 4\pi\varepsilon_0 r, \qquad (1)$$

where $q$ is the dust particle charge, $r$ is the distance between any two particles and $\lambda_D$ is the dust Debye length. Thus, a Yukawa system confined by a parabolic potential well on the horizontal direction and a much stronger parabolic potential well on the vertical direction can act as a reasonable analogue for such systems, allowing the study of the physics behind the resulting finite 2D plasma crystals.

A recent major topic of research on finite 2D plasma crystals is their structural symmetry. It is well known both theoretically and experimentally that 2D plasma crystals generally exhibit hexagonal structure. However, the influence created by the confining potential necessary for the formation of a finite crystal can sometimes result in concentric ring structures, which can compete for overall system symmetry with the normative hexagonal structure. A 3D analogue for this competition between ordered forms has recently been shown inside large three-dimensional (3D) laser-cooled ion clusters confined within either a Penning or Paul trap. The majority of these clusters exhibit shell-like structure with small BCC lattices seen to form within their interior as larger clusters are formed [7]. The critical size for the transition between these types of structure has been determined to be around N = $10^4$.

Although tantalizing, an obvious fundamental difference exists between plasma crystals and this type of ion system; in a plasma crystal the interaction between particles can no longer be described by a pure Coulomb potential, but must instead employ a Yukawa, or screened Coulomb potential with a screening length (Debye length) $\lambda_D$. The impact this has on the overall competition between system symmetry forms is one of the basic questions examined by this paper with the necessary data gathered by numerically examining finite 2D Plasma Crystals of 1000 particles for varying values of $\lambda_D$ and assuming a Yukawa potential interparticle interaction. The dependence of the resulting internal structure on the dimensionless Debye length λ (as defined in Section II) is examined both qualitatively and quantitatively.

A second major topic of research for finite 2D Plasma Crystals (clusters) are the oscillation modes produced via either thermal or forced perturbation. Horizontal oscillation modes, which involve particle motion on the horizontal direction, have recently been investigated intensely [5], [6].

Theoretically however, the total spectra for 2D clusters should consist of modes involving both horizontal and vertical particle motion. Although dust lattice wave modes created by vertical particle motion have been investigated for both one-dimensional (1D) [8]-[10] and 2D plasma crystals [11], [12], vertical oscillation modes within *finite* plasma crystals (clusters) have not yet been examined. Accordingly in this research, both vertical and horizontal oscillation modes are obtained for particle numbers between $N = 3$ and $N = 150$ again employing a box_tree simulation of thermally excited finite 2D plasma crystals. The resulting horizontal mode spectra is compared with previously published experimental and theoretical results while the vertical mode spectra is analyzed and compared with independent analytical results.

## II. NUMERICAL SIMULATION

In this research, the formation of finite 2D plasma crystals was simulated employing a box_tree code. The box_tree code used is a Barnes-Hut tree code written by Richardson [13] and later modified by Matthews and Hyde [14], Vasut and Hyde [15], and Qiao and Hyde [11], [12] to simulate complex plasmas systems under various conditions. Box_tree models systems composed of large numbers of particles by dividing the 3D box containing them into self-similar nested sub-boxes and then calculating all interparticle interactions using an incorporated tree code. Since the majority of these interparticle interactions can be determined by simply examining the multipole expansion of the collections of particles within the sub-boxes, the code scales as NlogN instead of $N^2$, resulting in much greater CPU time efficiency than is possible employing a traditional molecular dynamics approach.

As described above, the interparticle potential is modeled herein as a Yukawa potential of the form given by (1). The external confining potential is assumed to be parabolic in 3D,

$$E_{ext}(x, y, z) = \frac{m}{2}\left[\omega_{xy}^2(x^2 + y^2) + \omega_z^2 z^2\right] \quad (2)$$

where $x$, $y$, $z$ are representative particle coordinates and $\omega_{xy}^2$ and $\omega_z^2$ is the parabolic confinement magnitude on the horizontal and vertical directions respectively. Despite its simplicity, this potential model captures the basic properties of a multitude of classical systems and serves as an important reference point for more complex confinement situations [16].

Particles are contained in a 10×10×10 cm cubic box, which is large enough that particles, confined by the external potential, always remain within the box for the various Debye lengths examined. Particles are assumed to have constant and equal masses of $m_d = 1.74 \times 10^{-12}$ kg, equal charge of $q = 3.84 \times 10^{-16}$ C and equal radii of $r_0 = 6.5$ μm. Initially all simulations assume a random distribution of particles subject to the condition that the center of mass of the particle system must be located at the center of the box. Thermal equilibrium for the particle system at a specified temperature is established by allowing collisions of the dust particles with neutral gas particles. Collisions are assumed to be elastic with both momentum and kinetic energy conserved and $10^5$ collisions for each particle during each time step are allowed. Particles are initially given a zero velocity; they obtain higher velocities almost immediately after the start of the simulation due to potential interactions and then cool via collisions with neutrals. Crystal formation occurs less than 5 seconds into the simulation with the particle system continuing to cool slowly thereafter and reaching a thermal equilibrium in approximately 30 seconds. Results are not dependent on particle radius since particles are modeled as point masses where the radius as given above is used only for calculation of particle mass. Dimensionless lengths and energies are employed throughout by introducing the units $r_0 = (q^2/2\pi\varepsilon m\omega_{xy}^2)^{1/3}$ and $E_0 = (m\omega_{xy}^2 q^4/32\pi^2\varepsilon^2)^{1/3}$, respectively [16]. Accordingly a dimensionless Debye length, defined as $\lambda = \lambda_D/r_0$, will be used when investigating the relationship between the Debye length and the system's structural phases. It is important to note that with this change, the effect of the confining parameter $\omega_{xy}$ is now included within $\lambda$. Previously published results have shown that for zero temperature, the system's structural phase can be completely determined by the dimensionless Debye length, $\lambda$ [16], [17].

### III. CRYSTAL STRUCTURE

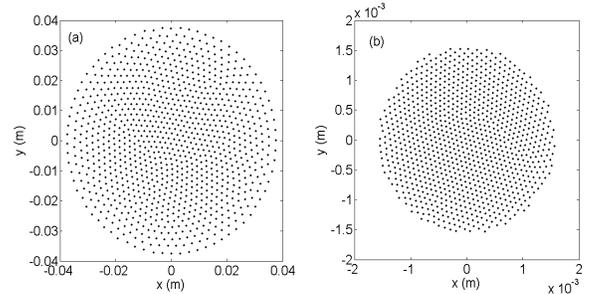

Fig.1. Top view of two finite 2D plasma crystals consisting of 1000 particles each with a (a) $\lambda$=7.3123 and (b) $\lambda$=.0024 at T = 10k.

For the initial phase of this research, systems having various Debye lengths ($\lambda$ ranging from .0024 to 7.3123) and consisting of 1000 particles were simulated at four different temperatures, 1000K, 100K, 10k and 1K. For large values of $\lambda$ (e.g. $\lambda$=7.3123) a 2D lattice with hexagonal interior symmetry and a few concentric rings along the outer edge is generally formed as can be seen in Fig. 1 (a). As the value of $\lambda$ decreases, these outer rings disappear with the overall system symmetry transitioning to a "pure" hexagonal lattice for very small values (e.g. $\lambda$=.0024) of $\lambda$ (Fig. 1 (b)). This process can be more clearly seen by examining the particle radial distribution function, as shown in Figures 2 (a) and (b) where $\lambda$=7.3123 and $\lambda$=.0024 and T=10K. At least 3 concentric rings can be clearly seen along the outer edge of the lattice for $\lambda$=7.3123 with no rings at all forming for $\lambda$=.0024. Figures 2 (c) and (d) show the corresponding pair correlation functions for these cases where it can be seen that for both large and small Debye length, lattice symmetry remains hexagonal in the interior.

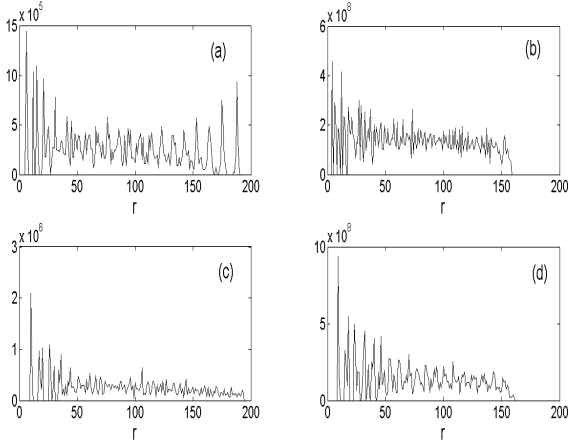

Fig.2. Radial density distribution functions for finite 2D plasma crystals with (a) $\lambda$=7.3123 and (b) $\lambda$=.0024. The corresponding pair correlation functions assuming the same values of $\lambda$ are shown in (c) and (d). (For all cases, N = 1000 and T = 10K.)

To quantitatively investigate the effect (if any) that $\lambda$ has on this transition, the potential energy of the system versus $\lambda$ was calculated for four different temperatures. As shown in Fig.3 (a) and as expected, it was determined that for identical Debye lengths, systems with higher temperature had larger potentials. Surprisingly however, the potential *was* found to be far more dependent on Debye length than on temperature. Fig.3 (a) shows the logarithmic response of the system potential versus $\lambda$. As can be seen, although the system potential does continue to increase across all values of $\lambda$, it is neither linear nor parabolic in nature. As shown, for $\lambda < .2437$, the increase in slope for E versus $\lambda$ is larger than when $\lambda > .2437$. This effect can be more clearly seen in Fig.3 (b), where the interaction potential $E_{int}$ is used instead of the total potential E and only the case for temperature T = 100k is considered. Here both E and $E_{int}$ are dimensionless energies normalized by the unit energy $E_0$ as discussed above. Under these conditions, the slopes for $\lambda < .2437$ and for $\lambda > .2437$ are found to be 1.678 and 0.652, respectively. Thus, two different types of structure are implied as evidenced by the dependence of the potential on $\lambda$ and the transition seen at $\lambda_{critical} \approx .2437$. As can be seen in Fig. 3(a), the dependence of the potential on the temperature is small when compared to its dependence on $\lambda$ allowing it to be neglected for this transition. In other words, it can be assumed that $\lambda_{critical}$ does not depend on temperature.

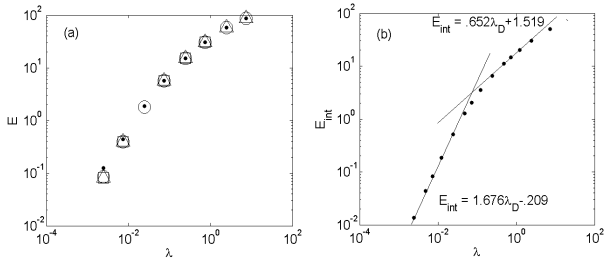

Fig.3. The total potential vs $\lambda$ for (a) various temperatures. (In the above, the dots represent T = 1000k, circles represent T = 100k, squares represent T = 10k, and triangles represent T = 1k.) (b) The interaction potential vs $\lambda$ for T =

100. (Both cases are for finite 2D plasma crystals consisting of 1000 particles.)

## IV. VERTICAL OSCILLATION MODES

For the second question to be addressed in this paper, the formation of finite 2D plasma crystals for particle numbers between $N = 3$ and $N = 150$ were examined. In all cases, the dimensionless Debye length was held constant at $\lambda = 0.1389$. The external potential within the sheath was assumed to be as given by (2) with $\omega_{xy}^2 = 2.21 s^{-2}$ and $\omega_z^2 = 221.07 s^{-2}$. Once plasma crystals reached thermal equilibrium, the thermal motion of the dust particles around their equilibrium positions was tracked for 32 seconds with output data files created every 0.04 second yielding a total of 800 data files. These data were then used to obtain the normal mode spectra employing the method given by Melzer [6], taking into account not only the horizontal but also the vertical motion of all cluster particles. Additionally, the normal mode spectra were calculated analytically using the equilibrium positions of the particles as defined by

$$E_{\alpha\beta,ij} = \frac{\partial^2 E}{\partial r_{\alpha,i} \partial r_{\beta,j}} \qquad (3)$$

with $\alpha$ and $\beta$ = x, y and i, j denoting the particle number.

The spectra of the resulting normal modes, including both the horizontal and vertical modes are given in Fig. 4 (d-f). As expected, the horizontal mode spectra is in agreement with previously published results [5], [6]. The vertical mode spectra are shown in Fig. 4 (g-i) where the maximum frequency for any given vertical mode is known as the vertical oscillation frequency. For the case at hand $\omega_z = 14.87 s^{-1}$, corresponding to the vertical oscillation frequency which would be produced if the entire particle system acted as a solid plane. As can be seen, the mode frequency decreases as the mode number $l$ decreases.

It is interesting to note that for small numbers of particles, i.e. $N < 47$, the horizontal and vertical spectra can appear as two separate branches with the horizontal mode spectra again retaining agreement with previous results. As the overall particle number increases ($N \geq 47$), the minimum frequency for the vertical mode decreases until it falls below the maximum frequency for the horizontal modes, and a merging of the two branches occurs. In either case when considered separately, the horizontal mode spectra remains in agreement with previously published results and the vertical mode frequency continues to decrease as the mode number decreases, with the maximum frequency remaining the vertical oscillation frequency, $\omega_z$.

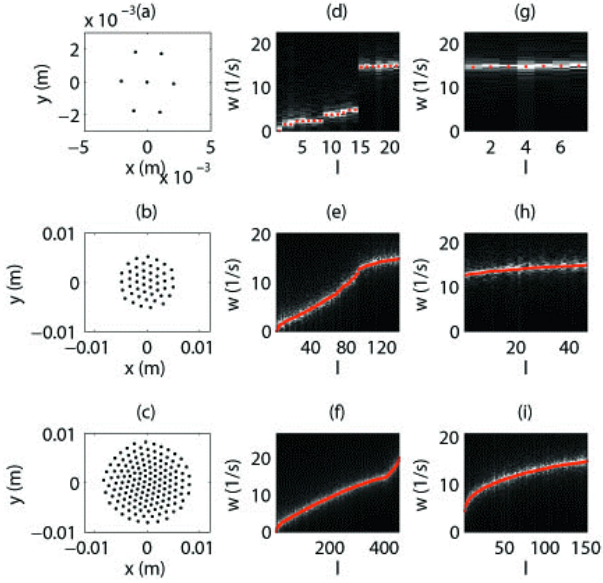

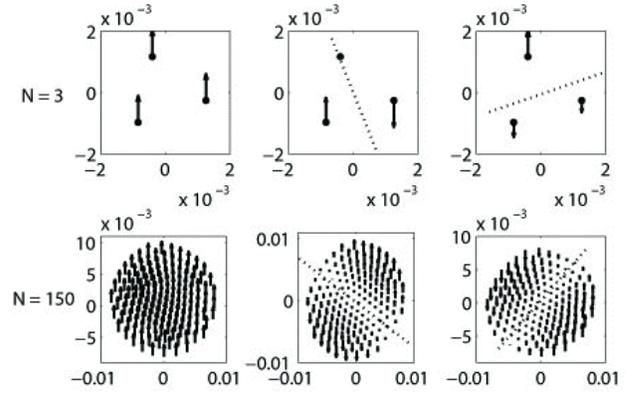

Fig.4. Initial cluster structure for (a) N = 6, (b) 46 and (c) 150. Corresponding mode spectra for all normal modes (d-f) and for vertical modes only (g-i) for (d, g) N = 6, (e, h) 46, and (f, i) 150.

A more thorough investigation of the vertical mode oscillation patterns can be conducted by examining the corresponding eigenvectors $\vec{e}_{i,l}$ and then determining the oscillation amplitude and direction for particle $i$ and mode number $l$. A detailed examination of this sort for the vertical modes obtained for the dust clusters mentioned above (particle number $3 < N < 150$) was conducted. As mentioned previously, the highest frequency mode was the mode corresponding to a vertical oscillation of the system of particles acting as a solid plane. This frequency was always equal to $\omega_z$ and independent of particle number. The modes with the second and third highest frequencies corresponded to rotational oscillations of the system (again as a solid plane) around differing horizontal axes. For an isotropic system, these modes are degenerate; in the system modeled above, they have slightly different frequencies due to the slight anisotropic nature of the cluster. For example, when $N = 150$, they have frequencies of $14.7935\text{s}^{-1}$ and $14.7934\text{s}^{-1}$ respectively. These three modes were found to exist for all clusters examined with representative data shown in Fig. 5 for the values of $N = 3$ and $N = 150$.

Fig.5. Representative oscillation patterns for the three highest frequency vertical modes for clusters having $N = 3$ and $N = 150$ as described in the text.

As the frequency and mode number decrease, the vertical modes begin to exhibit more complex oscillation patterns. This can be seen in Fig. 6 where the oscillation patterns for the next highest frequency modes in the vertical direction (after the three modes described above) are shown for $N = 150$. As can be seen, the oscillation pattern for a specific mode generally assumes the form of a Bessel function $J_m(k_{mn}r)$ in the radial direction $r$ and a Fourier function $e^{im\theta}$ in the angular direction $\theta$. (In this case, $m$ is equal to the number of periods in the angular direction while $n$ is determined by the boundary conditions on the radial direction.) As expected, boundary conditions at the cluster's edge are not well represented by closed boundary conditions, where the magnitude of the oscillation should be zero, or by free boundary conditions where the slope should be zero. In reality, the slope at the boundary generally assumes either a maximum or a minimum, i.e., the third derivative of the particle displacement is zero. Thus, if the first zero of the third derivative of the particle displacement is at the boundary, $n = 1$; if the second zero is at the boundary, $n = 2$ … and so on.

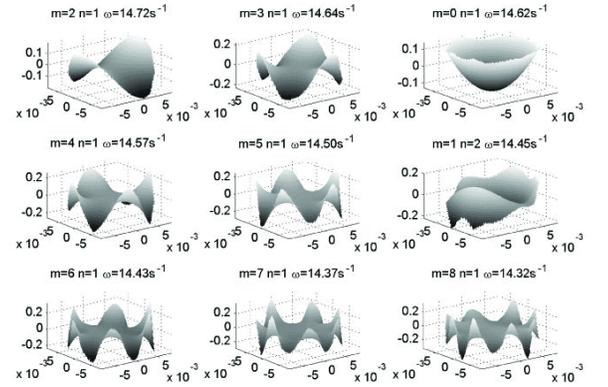

Fig.6. Representative oscillation patterns for vertical modes having frequencies $\omega_z \geq \omega \geq 14.32s^{-1}$ ($N = 150$).

At the other limit, as shown in Fig. 7, for the lowest frequencies (N = 150) peaks and valleys only appear at the center of the cluster while all other areas of the cluster remain flat. Thus, the maximum energy (amplitude) for the vertical motion of the particles remains concentrated across a few inner rings with the outer rings remaining almost motionless in the vertical direction. This is contrary to the manner in which the horizontal modes act where it is for the highest frequency modes that the particle motion is concentrated within the inner two rings and the outer rings remain more or less motionless. Taken together, for a 2D dust cluster with large particle numbers, the horizontal particle motion is distributed throughout the cluster while the vertical motion remains concentrated at the cluster's center for low frequencies with exactly the opposite occurring for high frequencies.

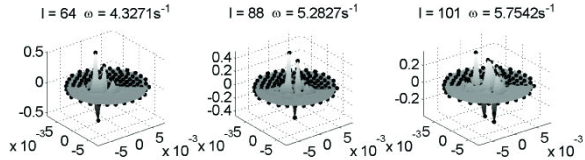

Fig.7. Representative oscillation patterns for the vertical modes having the lowest frequencies ($N = 150$).

## V. CONCLUSION

In this research, the formation of finite 2D plasma crystals was numerically simulated employing a box_tree code and assuming a Yukawa system confined on both the vertical and horizontal directions. Particle systems consisting of 1000 particles and having Debye lengths $\lambda$ ranging from .0024 to 7.3123 were simulated at four different temperatures, 1000K, 100K, 10k and 1K. It was found that for large values of $\lambda$, 2D lattices showed hexagonal symmetry in their interior with a small number of concentric rings forming along their outer edge while for very small values of $\lambda$ "pure" hexagonal lattices were formed. This tendency was confirmed by comparing data collected from the lattice view with both the radial density distribution function and the pair correlation function.

The potential energy of the system was also calculated as a function of Debye length for various temperatures. It was shown that the transition between two structural types can be clearly seen by plotting the interaction potential as a function of the Debye length and that a critical Debye length $\lambda_{critical} \approx .2437$ can be identified. It was also shown that for identical Debye lengths, systems with higher temperature had larger potentials, as expected.

In the second phase of the study, vertical and horizontal oscillation modes for thermally excited 2D dust coulomb clusters were investigated. Both a box_tree numerical simulation and an analytical method were employed to provide data for clusters with particle numbers ranging between $N = 3$ and $N = 150$. The resulting horizontal mode spectra were shown to agree with previously published results while the vertical mode spectra obtained from the box_tree simulation were compared with analytical results and shown to agree self-consistently. The maximum frequency ($\omega_z$) for vertical mode oscillations was found to be the same as the vertical oscillation frequency for the system acting as a solid plane with the mode frequency decreasing as the mode number $l$ decreases. For clusters with small particle numbers ($N < 47$), all vertical modes show higher frequencies than corresponding horizontal modes. For larger clusters ($N \geq 47$), the vertical and horizontal modes are co-mingled.

Additionally, the resulting oscillation patterns for each of the vertical modes have been investigated. It was found that three highest frequency modes exist for all clusters examined ($3 < N < 150$). The highest frequency mode is the mode corresponding to a vertical oscillation of the entire system of particles as a solid plane. This frequency is always equal to $\omega_z$ and is independent of particle number. The modes with the second and third highest frequencies are quasi-degenerate and correspond to rotational oscillations of the system (again as a solid plane) around two different horizontal axes.

For clusters with large numbers of particles (N = 150, for example), the highest frequency modes after the three modes described above show oscillation patterns similar in shape to Bessel-Fourier functions with various indices of $m$ and $n$. At the other limit for the lowest frequency modes, the strongest particle motion within the cluster is concentrated within the first few inner rings with the outermost rings remaining almost motionless. In contrast, the horizontal modes show the strongest particle motion concentrated within the inner rings at their highest frequencies.

2D complex plasma crystals with inner ring hexagonal symmetry and concentric rings along their outer edge as described in this research have been observed experimentally. (For one example, please see [18]). Wave modes created by vertical particle motion as considered in the second part of this paper have also been examined ([9], [10]). However, the detailed, quantitative results predicted in this paper have not yet been verified by experiments. Current experiments to verify the above are underway within CASPER.

**Ke Qiao** received the B.S. degree in physics from Shandong University, Jinan, China, and the Ph.D. degree in theoretical physics from Baylor University, Waco, TX. He is currently with Baylor University, where he is an Assistant Research Scientist at the Center for Astrophysics, Space Physics, and Engineering Research (CASPER). His research interests include structure analysis, waves and instabilities, and phase transitions in complex (dusty) plasmas.

**Truell W. Hyde** (M'01) received the B.S. degree in physics from Southern Nazarene University, Bethany, OK, and the Ph.D. degree in theoretical physics from Baylor University, Waco, TX. He is currently with Baylor University, where he is the Director of the Center for Astrophysics, Space Physics and Engineering Research, a Professor of physics, and the Vice Provost for Research in the university. His research interests include space physics, shock physics and waves, and nonlinear phenomena in complex (dusty) plasmas.